\documentclass[pdftex,twocolumn,10pt,letterpaper]{extarticle}

\newcommand{\AUTHORS}{Hyeontaek Lim, Vyas Sekar, Yoshihisa Abe, David G. Andersen}
\newcommand{\NAME}{NetMemex\xspace}
\newcommand{\TITLE}{{NetMemex}: Providing Full-Fidelity Traffic Archival}
\newcommand{\KEYWORDS}{}
\newcommand{\CONFERENCE}{}
\newcommand{\PAGENUMBERS}{yes}       \newcommand{\COLOR}{yes}
\newcommand{\showComments}{yes}
\newcommand{\comment}[1]{}
\newcommand{\onlyAbstract}{no}

\usepackage{ifthen}
\ifthenelse{\equal{\PAGENUMBERS}{yes}}{\usepackage[nohead,
            left=0.76in,right=0.76in,top=1in,
            footskip=0.5in,bottom=1in,                 columnsep=0.33in
            ]{geometry}
}{\usepackage[noheadfoot,
			left=0.76in,right=0.76in,top=1in,
            footskip=0.5in,bottom=1in,
            columnsep=0.33in
	    ]{geometry}
}

\usepackage[font=bf]{caption}

\usepackage{titlesec}
\titleformat{\section}  {\bf\large\uppercase}{\thesection.\quad}{0pt}{}

\usepackage{enumitem}
\setlist{itemsep=0pt,parsep=0pt}             
\usepackage{fancyhdr}

\ifthenelse{\equal{\PAGENUMBERS}{yes}}{  \pagestyle{plain}
}{  \pagestyle{empty}
}

\usepackage[numbers,sort]{natbib}

\usepackage{booktabs}
\usepackage{color}
\usepackage{colortbl}
\usepackage{float}                                                                        
\usepackage{mathptmx}                        \usepackage{courier}
\usepackage[scaled=0.92]{helvet}

\ifthenelse{\equal{\COLOR}{yes}}{  \usepackage[colorlinks]{hyperref}}{  \usepackage[pdfborder={0 0 0}]{hyperref}}
\usepackage{url}

\hypersetup{pdfauthor = {\AUTHORS},
pdftitle = {\TITLE},
pdfsubject = {\CONFERENCE},
pdfkeywords = {\KEYWORDS},
bookmarksopen = {true}
}

\definecolor{placeholderbg}{rgb}{0.85,0.85,0.85}

\usepackage[pdftex]{graphicx}
\usepackage{soul}

\usepackage{textcomp}
\usepackage{multirow}

\usepackage{fixltx2e}

\usepackage{xspace}

\usepackage{balance}

\usepackage[absolute]{textpos}

\newfloat{acmcr}{b}{acmcr}

\newcommand{\note}[2]{
    \ifthenelse{\equal{\showComments}{yes}}{\textcolor{#1}{#2}}{}
}

\date{}
\title{\textbf{\TITLE}}
\author{{\large Hyeontaek Lim, Vyas Sekar, Yoshihisa Abe, David G. Andersen}\\
{\em Carnegie Mellon University}}

\usepackage{microtype}  
\begin{document}

\maketitle

\makeatletter{}\section*{ABSTRACT}

NetMemex explores efficient network traffic archival without any loss of information.
Unlike NetFlow-like aggregation, NetMemex allows retrieving the entire packet data including full payload,
which makes it useful in forensic analysis, networked and distributed system research, and network administration.
Different from packet trace dumps, NetMemex performs sophisticated data compression for small storage space use and optimizes the data layout for fast query processing.
NetMemex takes advantage of high-speed random access of flash drives and inexpensive storage space of hard disk drives.
These efforts lead to a cost-effective yet high-performance full traffic archival system.
We demonstrate that NetMemex can record full-fidelity traffic at near-Gbps rates using a single commodity machine, handling common queries at up to 90.1 K queries/second, at a low storage cost comparable to conventional hard disk-only traffic archival solutions.

\ifthenelse{\equal{\onlyAbstract}{no}}{\makeatletter{}\section{Introduction}
\label{sec:intro}

Archiving network traffic without loss of information is a useful foundation for activities including
forensic analysis~\cite{Moore2001,maier:sigcomm2008,www-snifferanalysis,Geambasu:netdb2007,www-networktimemachine,Desnoyers2007},
scientific research, particularly about networking and distributed systems~\cite{spring2000,Anand:sigcomm2009,Cheng:sigcomm2006},
network administration~\cite{www-endaceprobe,www-snifferanalysis},
and others.

However, the task of traffic archival is difficult, as recent applications impose technical challenges:
\begin{itemize}
	\item \textbf{Data volume}.
	Applications often require recording a few or more days of full traffic for analysis (e.g., forensic analysis)~\cite{maier:sigcomm2008,Papadogiannakis:mascots2010},
	resulting in a large amount of data to store ($10.8+$~TB/Gbps/day without compression).
	\item \textbf{Recording performance}.
	It has become more common to record high-speed traffic (e.g., $1$~Gbps rate)~\cite{maier:sigcomm2008,Anderson2006b,Degioanni:imc2004,www-networktimemachine}.
	\item \textbf{Query performance}.
	There is an increasing demand for fast query processing~\cite{maier:sigcomm2008,Fusco:2010vldbendow,Fusco:ccr2012}.
\end{itemize}

Previous solutions fail to fully address these challenges:

(1) \emph{Lossy traffic archival}.
There are numerous works that mitigate high storage consumption by aggregating or selectively recording traffic data~\cite{maier:sigcomm2008,Cranor:sigmod2003,Papadogiannakis:mascots2010}.
However, this lossy approach is inadequate for applications that cannot tolerate any loss of information from the original traffic.
A few compression schemes have been proposed for compressing traffic information~\cite{Fusco:imc2012}, but they focus only on well-structured formats such as NetFlow records.

(2) \emph{Imbalance between recording and query performance, storage use}.
Many traffic archival systems have difficulty in achieving both high recording and query processing performance.
One of the most widespread methods is to focus on recording throughput by viewing hard disk drives as a huge circular queue and streaming packets to the queue as they arrive~\cite{www-endaceprobe,Desnoyers2007,maier:sigcomm2008,Fusco:ccr2012}.
In this technique, query processing can waste a large amount of I/O bandwidth due to high false positive rates incurred by inefficient indexing over such an unstructured data layout.
Some systems aim instead for high query performance~\cite{Geambasu:netdb2007}, but they do not target high recording speed, so these approaches are only viable for offline or semi-online recording.
These works do not perform detailed evaluation on the total storage cost, including the traffic data and any other metadata such as indexes, unless their system design focuses on reducing the storage space use~\cite{maier:sigcomm2008,Papadogiannakis:mascots2010}.

In this paper, we present \NAME, a full-fidelity traffic archival system that simultaneously attains high recording and query performance and efficient storage.
A single-node server-class system running \NAME can archive Internet traffic of $0.62$--$1.15$ Gbps, and it can handle up to $90.1$ K common queries per second.
While \NAME employs flash drives, which are expensive relative to hard disk drives, because of \NAME's data compression techniques, its total storage cost is similar to conventional hard disk only solutions.

The three key elements of our approach are:
\begin{itemize}
\item \textbf{Flow-oriented data reorganization and indexing} increases the efficiency of query processing and the effectiveness of data compression by transforming the packet stream into a flow-oriented form;
\item \textbf{Employing both flash and hard disk} improves query latency by storing frequently and randomly accessed small data on flash while keeping less frequently accessed bulk data on hard disk; and
\item \textbf{Extensive data compression} reduces the total storage cost by using effective data compression techniques, which are applied to the packet header and payload separately.
\end{itemize}

These approaches are not a combination of piecewise techniques; they work together to help \NAME achieve its performance and cost goals.
For example, flow reordering substantially increases the effectiveness of compression,
and data compression enables storing frequently accessed data on flash without a significant increase in the total storage cost.

\makeatletter{}\section{Background}
\label{sec:background}

We first describe a general traffic archiving system and its challenges, and then we show several data compression techniques and storage technologies that we use in \NAME.

\subsection{Traffic Archival Systems}

A traffic archival system is a storage system specialized for network traffic data storage and retrieval.
This system stores network packets and returns the stored information upon request.

\begin{figure}
\centering
\includegraphics[width=0.25\textwidth]{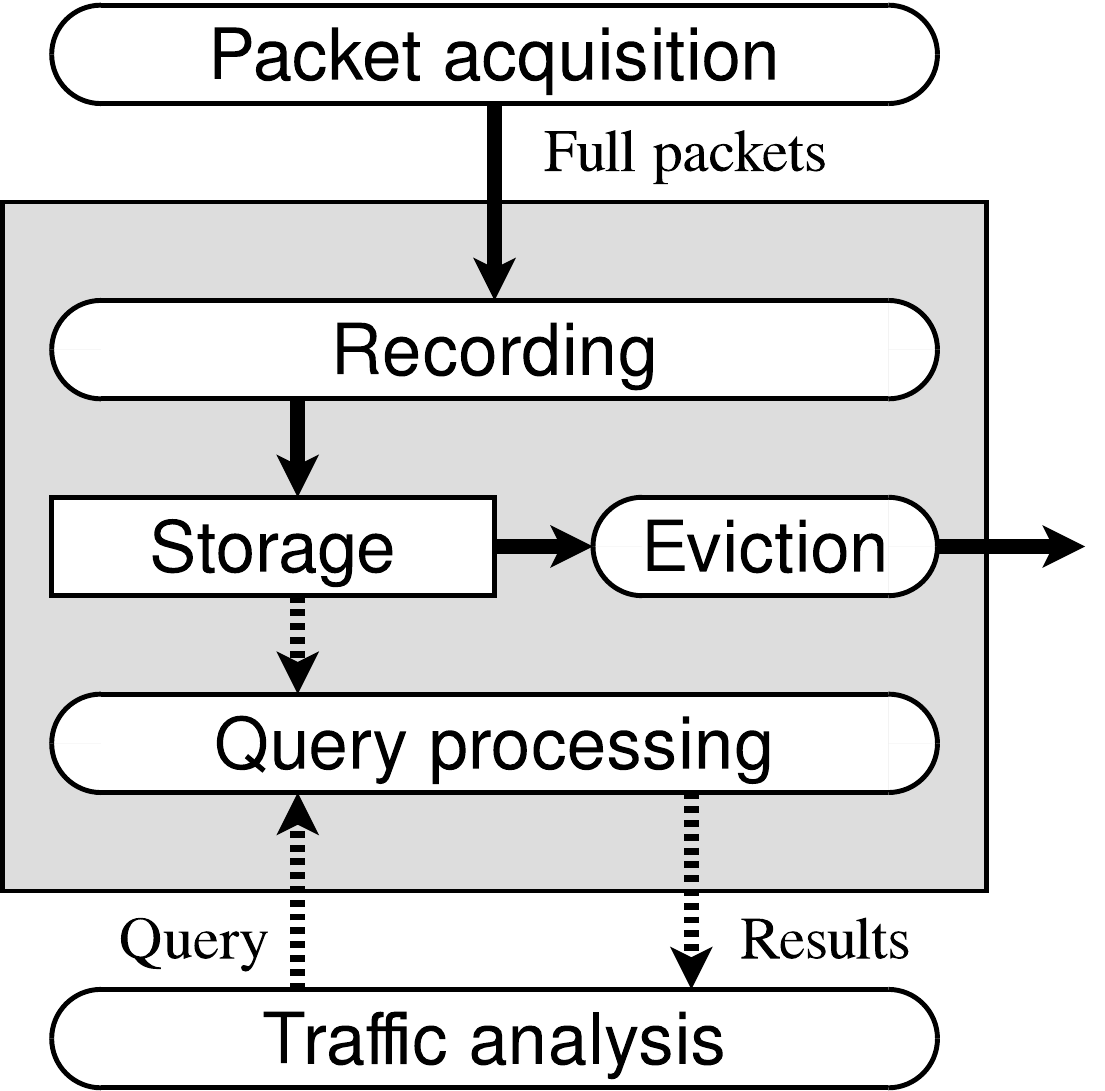}
\caption{
	High-level view of a traffic archiving system (within the gray box) and its interaction with other system components (outside of the gray box).
	Solid arrows show data processing, storage, and eviction, and dashed arrows are data retrieval.
}
\label{fig:overview}
\end{figure}

Figure~\ref{fig:overview} shows typical components of a traffic archiving system.
A traffic archiving system has a simple interface.
The system receives captured packets from an external packet acquisition method, such as tcpdump/libpcap~\cite{www-tcpdump} and Endace DAG~\cite{www-endaceprobe}, and keeps them for a specific period of time.
Stored packets and traffic information can be requested by traffic analysis software.

Inside the system, incoming packets go through the stages of recording and indexing, storage, and eviction.
The recording and indexing phase accepts the input packets and stores them to storage.
The packets can be transformed in this stage if there is a better data format available. As the storage fills, the system evicts old data to make room for new packets.
The evicted data can be either discarded or stored in other storage for later analysis and longer archival; the auxiliary storage can be a larger-scale traffic archiving system or a large-scale ordinary storage cluster.

\subsection{Data Compression}

\textbf{Dictionary-based compression} techniques are most commonly used in everyday data compression.
They maintain recently seen data patterns in a data structure called a dictionary.
They use the dictionary to find a duplicate pattern in the new data,
and they replace the pattern with a reference to the pattern in the dictionary.
Since the reference can be encoded in smaller space than the original pattern in many cases,
they can reduce the total space required to represent the original data.

A larger dictionary often (but not always) allows a higher compression ratio; unfortunately, it also typically decreases compression speed due to more expensive dictionary search operation.
Therefore, fast compression techniques use small dictionary sizes (e.g., 32 KiB for zlib~\cite{www-zlib}).
Redundancy elimination~\cite{spring2000,Aggarwal2010}, a form of dictionary-based compression on network routers, also maintain a small time window to operate at a high rate.

\textbf{Data deduplication} techniques~\cite{Lilley00},
unlike dictionary-based compression, 
efficiently detect and reduce
data redundancy across longer ranges. 
They use ``fingerprints'' to divide data stream into smaller blocks called \emph{chunks}.
Each chunk is hashed using a strong hash function (e.g., SHA-1) to generate a unique identifier,
and they store only one copy of chunks with the same hash.
Because it is possible to maintain an index of chunk hashes that are collected for a large amount of data (e.g., TBs) and a long period of time (e.g., days),
data deduplication can save space even if redundant data are scattered across the entire data.

Dictionary-based compression and data deduplication are often used together to increase the compression effectiveness; each chunk is compressed using a dictionary-based compression technique
while chunks are deduplicated using a data deduplication technique.

\textbf{Header compression} techniques~\cite{rfc1144,rfc2507,Lilley00}
reduce the size of packet headers by exploiting redundancy in packets in the same flow (i.e., between packets sharing the same 5-tuple).
For example, the sequence and acknowledge numbers in a TCP flow have small increments, which can be encoded in one or two octets instead of the full four octets.
Header compression is typically applied between network links or end-points to reduce the bandwidth required to transit packets.

\subsection{Storage Systems and Technologies}

\textbf{Log-structured file systems}~\cite{rosenblum:lfs1992,kawaguchi:usenix1995}
convert random writes to sequential writes by appending new data to a log.
This approach takes advantage of the fact that many storage devices are good at sequential writes.

\textbf{Key-value stores}~\cite{decania:sosp2007,Andersen:fawn-sosp2009,www-memcached}
provide a high-performance storage service with a hash table-like interface, e.g., \texttt{GET(key)} and \texttt{PUT(key, value)}.
They are typically optimized to handle a large number of small items efficiently in space, time, or both.
A key-value store is a useful building block for data deduplication;
the system can use a chunk hash as a key to check if the chunk has been seen by looking its hash.

\textbf{Flash and hard disk} devices are commonly used storage devices nowadays.
Flash is a NAND-based solid-state storage technology that provides high random read speed, often exceeding 35 K small reads per second~\cite{www-intel-ssd-x25m};
however, its random write speed is slow (e.g., 300 small writes per second);
flash cannot perform in-place updates (a large ``erase block'' must be erased before being written) and typically uses a log structure internally, which is still unable to provide high random write throughput for small writes.

Hard disks are mechanical storage devices; it uses a spinning disk and moving I/O head.
Due to physical limitations, hard disks are bad at random access.
On the other hand, they offer fast sequential read and write and provide inexpensive storage space; as of 2013, a hard disk drive is $15.85$ times cheaper than a flash drive in terms of \$ per GB\@.

\makeatletter{}\section{Design}
\label{sec:design}

In this section, we first provide an overview of the design of \NAME,
and then we describe each component of \NAME in detail.

\subsection{Architecture}
\label{sec:architecture}

\begin{figure}
	\centering
	\includegraphics[width=0.45\textwidth]{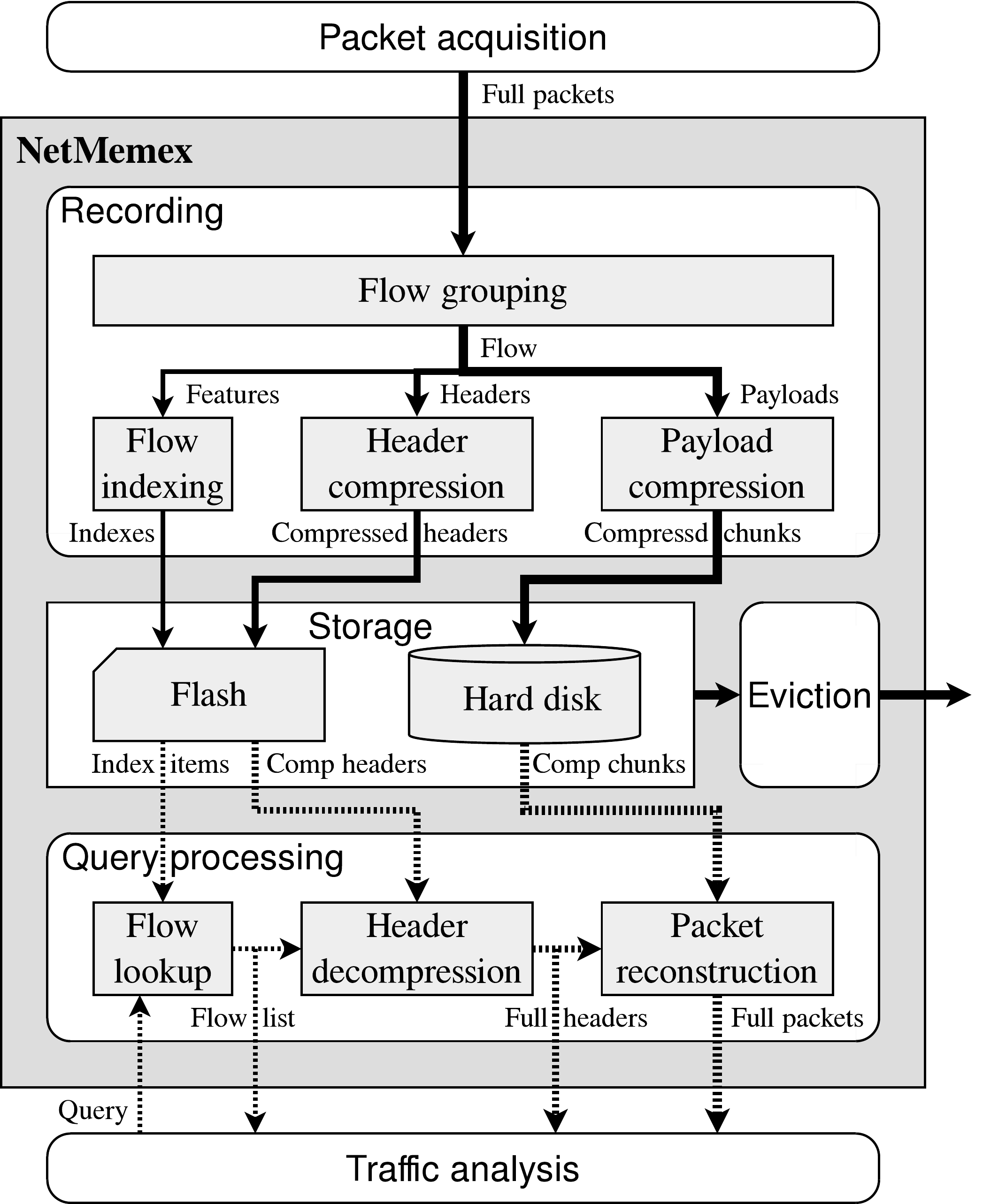}
	\caption{Architecture of \NAME.}
	\label{fig:architecture}
\end{figure}

Figure~\ref{fig:architecture} depicts the architecture of \NAME.
Within the basic structure of a traffic archiving system,
\NAME combines multiple techniques to manipulate network traffic data.

Packets are acquired using packet capture software and transferred to \NAME in the standard libpcap format.

\NAME records the full packets by transforming them into a compressed form that can be selectively accessed.
\NAME first groups the packets into flows (\S\ref{sec:flow}).
It applies header compression (\S\ref{sec:header}) and payload compression (\S\ref{sec:payload}) to the headers and payloads of each flow, followed by flow indexing (\S\ref{sec:indexing}).
Through this process, \NAME makes traffic data smaller and easy for query processing (\S\ref{sec:queryproc}), where each step preserves the full fidelity of the original traffic.

\NAME stores data using both flash drives and hard disk drives, to take advantage of the strength of each storage class;
flash holds headers and indexes, which are small and frequently accessed, and disk holds bulk payload data on disk.

On both storage device types, \NAME maintains data in a log-structured way in order to handle recording and eviction efficiently (\S\ref{sec:eviction}), and to extend the lifetime of the flash drives.

To enable high-speed query processing, \NAME allows an analysis application to choose how much detail about the original traffic is necessary.
For fastest operation, the application can ask \NAME if any matching flow exists.
It can retrieve a full list of matching flows if necessary.
If packet headers matter, but high query performance is still needed, \NAME can return full headers from flash drives.
When the full fidelity is necessary, the application can request full packet reconstruction, which will make \NAME fetch payload data from hard disk and construct a view of the original traffic.

Query processing produces full headers and packets in the libpcap format so that existing traffic analysis applications can easily use query results.

\subsection{Flow Grouping and Lifetime}
\label{sec:flow}

A \emph{flow} in \NAME is a sequence of packets with the same 5-tuple (source and destination IP address, protocol, source and destination port number if exist), ordered by arrival time (packet timestamp).
\emph{Flow grouping} is the first recording step that transforms the packet stream into a query-friendly form and facilitates per-flow compression in the subsequent recording process by grouping input packets into flows.

\begin{figure}
\centering
\includegraphics[width=0.45\textwidth]{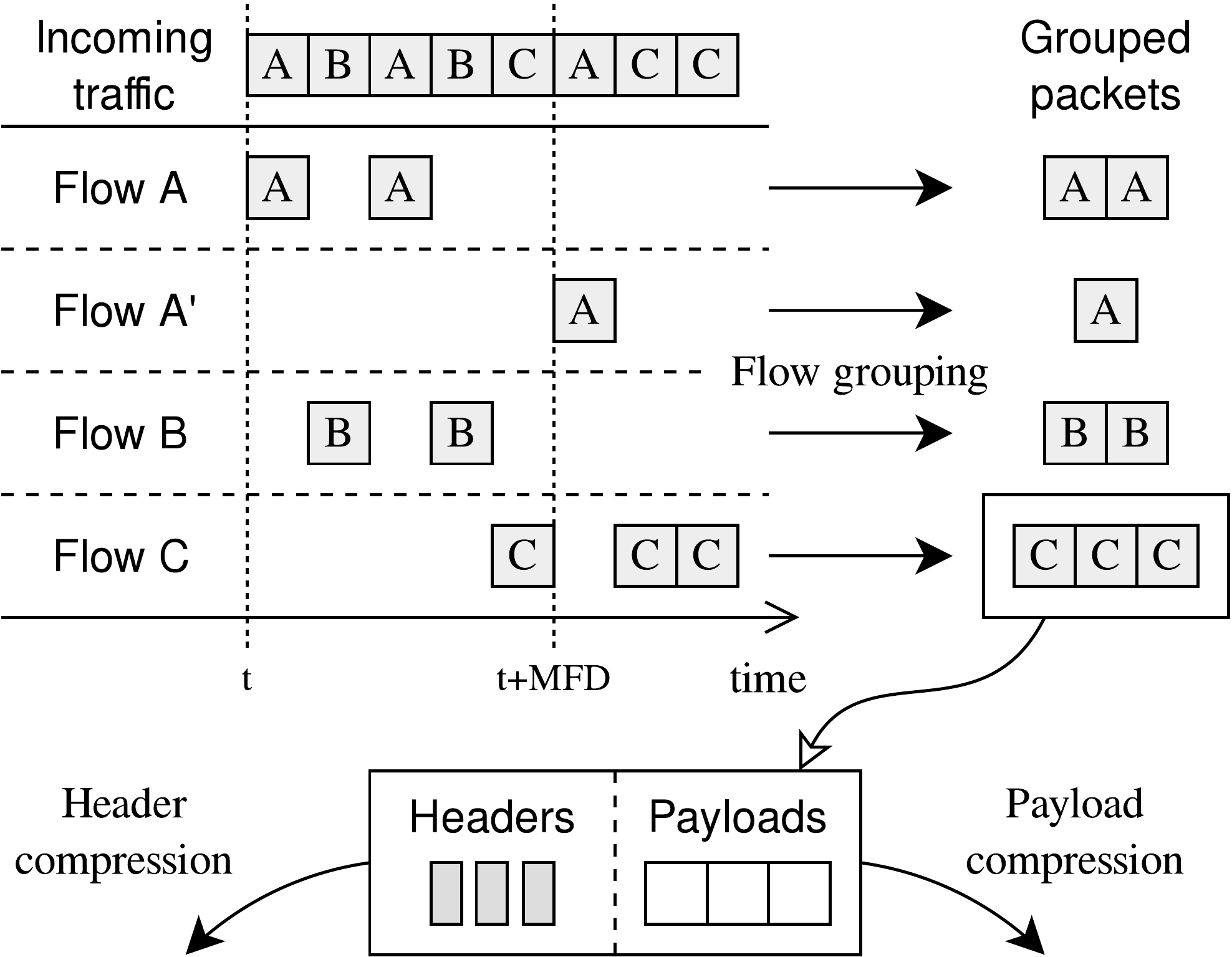}
\caption{
	Example of flow grouping in \NAME.
	The 5-tuple ID of each packet are labeled in the square.
	The packets of the 5-tuple~\texttt{A} are grouped into two flows
	because the lifetime of any flow is disallowed from exceeding the maximum flow duration (MFD).
	Packets of each flow are separated into headers and payloads for the subsequent processing of header compression and payload compression.
}
\label{fig:flowgrouping}
\end{figure}

Figure~\ref{fig:flowgrouping} illustrates how flow grouping works.
\NAME classifies incoming packets according to their 5-tuple and enqueues them to corresponding buffers.
When a 5-tuple stream terminates, \NAME flushes all packets from the buffer of the 5-tuple, and these packets form a flow.
It sends completed flows to the next recording stage to compress and index them.

\NAME imposes a \emph{maximum flow duration} (MFD) on each flow to bound the memory consumption of packet buffering.
Without limits on flow duration, flow grouping could demand an unbounded amount of memory for buffers when a flow runs for a long period.
To solve this problem, \NAME stops adding new packets to a flow if its lifetime is about to exceed the MFD.
In Figure~\ref{fig:flowgrouping}, for instance, 5-tuple~\texttt{A} actually generates two flows in \NAME, since the flow duration would exceed if all packets are grouped into a single flow.
As a consequence, \NAME can regulate the total buffer size to be within $\mbox{(input traffic rate)} \times \mbox{MFD}$,
and the memory requirement of flow grouping becomes more predictable and manageable.

Flow grouping does not destroy any of the traffic information because each packet holds the timestamp.
When required, analysis programs can sort packets by recorded timestamps to obtain the original packet arrival order.

\subsection{Header Compression}
\label{sec:header}

In this paper, a \emph{header} refers to a packet header and per-packet metadata (i.e., timestamp, packet lengths).
Headers contain key information for queries, such as hosts, port numbers, and TCP sequence numbers.

\NAME applies a variant of \emph{header compression}~\cite{rfc1144,rfc2507} to packet headers in order to efficiently reduce their size.
Conventional compression algorithms, such as zlib~\cite{www-zlib} and LZO~\cite{www-lzo}, aim to find and compress exact byte string matches;
unfortunately, these algorithms do not work well directly for headers,
as headers are a formatted data structure consisting of short and varying data fields.
Instead, header compression typically encodes differences between two consecutive headers instead of recording the full header content.

To maximize the effectiveness of header compression,
it should be applied to a flow consisting of packets with the same 5-tuple.
This is seamlessly done in \NAME because of the flow grouping step.

\begin{figure*}
	\centering
	\includegraphics[width=0.94\textwidth]{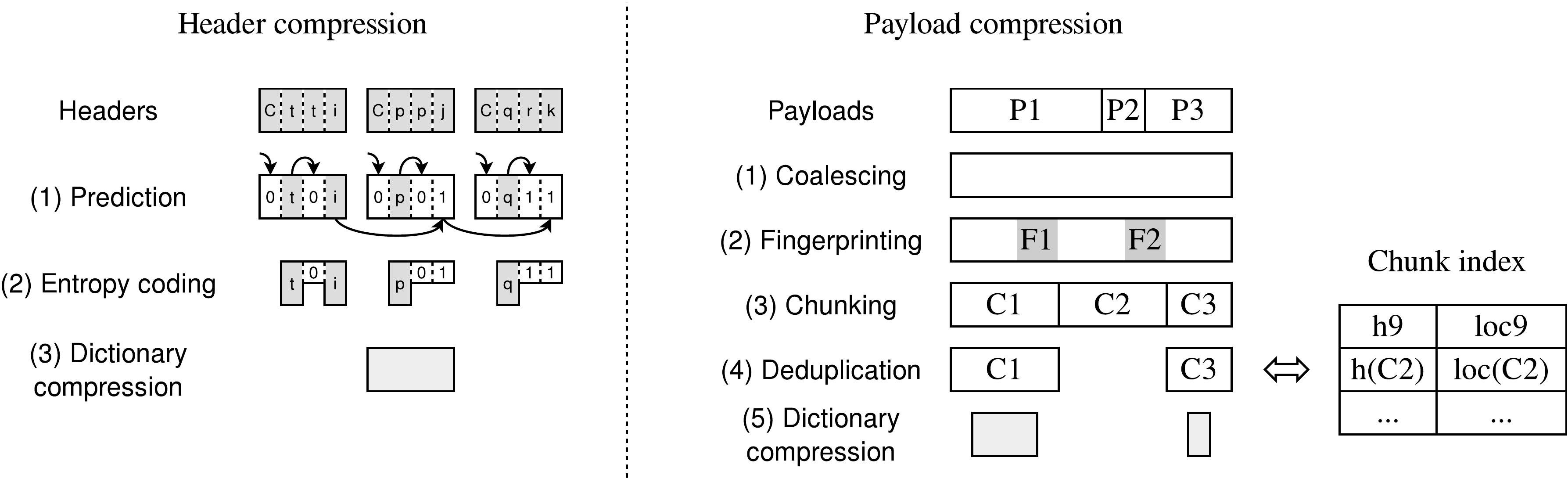}
	\caption{Schematic diagram of header compression and payload compression showing how a flow with three packets is compressed.}
	\label{fig:compression}
\end{figure*}

Compared to the standard header compression, \NAME's header compression (Figure~\ref{fig:compression}) is unique in two aspects:

\NAME applies \emph{intra-packet} compression extensively, in addition to \emph{intra-flow} compression of the conventional header compression.
\NAME can predict header field values using the other information within the same packet and store only differences between the predicted values and the actual values in the header.
For example, the IP packet length is typically inferable from the total packet length specified in the packet capture metadata.
The IP checksum is also a field that can be often omitted for storage when the checksum is valid;
if the checksums do not match, \NAME simply stores the invalid checksum found in the header to preserve the full fidelity.
This intra-packet compression is possible in \NAME but not typically in the general header compression due to possible transmit errors;
\NAME do not need to worry about errors as \NAME does not send compressed headers over network.

Because \NAME sees multiple packets in the same flow,
\NAME can use conventional dictionary compression as the final step of header compression.
This allows removing further redundancy;
for instance, when TCP sequence numbers simply increase by a constant,
the standard header compression would result in the same difference value for each packet;
however, by applying dictionary compression, \NAME can compress such repetition.

One (often negligible) side effect of header compression is that compressed header data becomes \emph{flow-addressable}, not packet-addressable.
One header can be read only after the decompression of the preceding header.
Hence, each flow is randomly accessible efficiently, but each packet in a flow requires partial or full decompression of the flow.
Since many queries request the first part of a flow or the whole flow rather than accessing random packets within it, this flow-addressability typically does not decrease query processing throughput.

After header compression, \NAME writes the data to flash storage.
The new data is appended to the most recently written one.
\NAME can easily delete the oldest headers by treating the storage as a circular queue.

\subsection{Payload Compression}
\label{sec:payload}

In packet payloads, \NAME takes a different strategy to reduce their volume because of their redundancy characteristics.
First, the internal structure of packet payloads is more irregular than headers.
Header compression largely relies on well-defined relations between packets.
However, many transport layer protocols, such as TCP, break this type of redundancy in the payload data.
For instance, they can adjust packet boundaries for various reasons (e.g., MSS, congestion window).
This adjustment shifts and cuts the byte stream of the content, making the redundancy between two packets less explicit.
Second, payloads frequently show abundant redundancy across different flows, apart from each other in time.
For example, two separate connections may download the same file with a large time gap (e.g., a few hours).
Because header compression is applied to each flow, it is less effective for these kinds of redundancy in the payload.

Instead, as illustrated in Figure~\ref{fig:compression}, \NAME compresses payloads using variable-length chunk deduplication, similar to LBFS~\cite{Muthitacharoen:sosp2001}, in conjunction with per-chunk dictionary compression.
It begins by constructing a byte stream by coalescing all the payloads in a flow,
which essentially ignores packet boundaries.
Then, chunking is done by scanning the stream and detecting chunk boundaries using fingerprinting~\cite{Rabin1981};
once chunks are discovered, each chunk is hashed using a collision-resistant hash function (e.g., SHA-1).
These chunk hashes act as keys when \NAME queries the \emph{chunk index}, an external key-value store, to determine whether each chunk is duplicate.
Finally, \NAME stores the location of the chunks together with the compressed headers.

The chunk index should be high-performance and cost-effective.
For a typical average chunk size of $4$ KiB, $1$ Gbps-rate traffic generates $30.5$ K chunks per second; each chunk will incur one lookup for the index,
so the chunk index must be fast to avoid stalling the recording process.
In addition, although payload compression may reduce the volume of the payload data, this savings can be compromised if the cost of maintaining the chunk index is high.
Even with the small chunk descriptor size (20-byte chunk hash and 8-byte chunk location for each chunk),
the chunk index is often too large to fit entirely in main memory (e.g., $73.8$ GB for one day for the above setting).

The chunk index keeps track of recent chunks only for a certain duration (\emph{deduplication window}).
If a chunk is older than this duration, its entry is removed from the chunk index (but this does not remove the chunk itself from disk); this mechanism controls how much system resources \NAME should use for deduplication.

The above performance and space requirements make a flash-based key-value store a good candidate for the chunk index.
We use SILT~\cite{Lim:sosp2011} as the chunk index in \NAME;
it provides fast operation using flash and has high space efficiency,
satisfying \NAME's requirements.

\subsection{Indexing}
\label{sec:indexing}

Once a flow is compressed, \NAME indexes it for accelerated query processing.
\NAME prepares indexes for common criteria such as IP addresses, port numbers, and so on;
whenever a query includes any of such criteria,
\NAME queries the index to significantly reduce the amount of data to read and inspect.

\begin{figure*}
\centering
\includegraphics[width=0.94\textwidth]{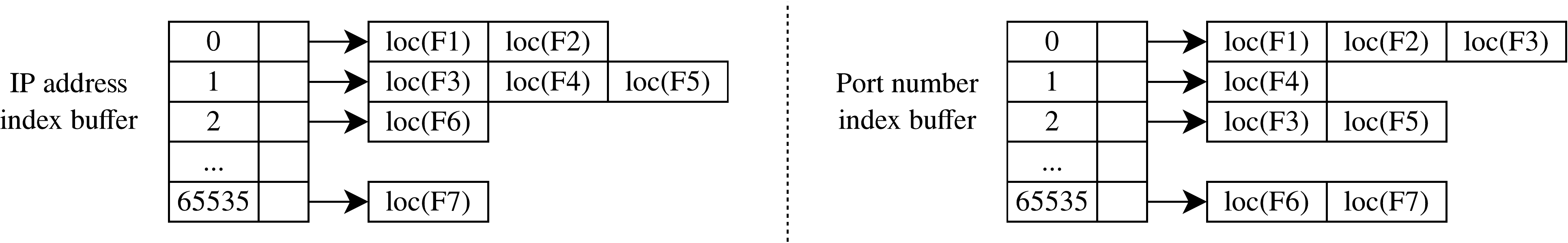}
\caption{
	Lightweight flow indexing for the IP address and the port number.
	Each bucket contains the location of flows whose address or port number is hashed into that bucket number.
}
\label{fig:index_buffer}
\end{figure*}

For lightweight indexing, \NAME's index is (1) flow-oriented, (2) compressed, (3) allows false positives, and (4) per-epoch.

\NAME indexes flows, not individual packets.
As shown in Figure~\ref{fig:index_buffer},
\NAME builds a hash table with $65,536$ buckets;
it hashes the indexed field value (e.g., IP address) to determine a bucket,
and inserts the location of the flow to the bucket.
Since all packets in the flow are stored consecutively by header compression,
\NAME can read the packets using the flow location only.

\NAME compresses these flow locations in each bucket using a similar way to header compression.
The location numbers are monotonely increasing after sort,
and thus \NAME can store the differences between two consecutive locations.
After compression, the index requires only $6.07$ bytes per flow per field.

However, there can exist hash collisions from using a small number for buckets,
and this leads to false positive answers (but no true negatives).
\NAME ensures that the output contains only true positive results,
by decompressing the packet headers and checking if the query criteria are met.

Finally, \NAME generates a set of independent indexes for each \emph{epoch}.
The indexes for an epoch describe the flows that have ended during that period of time.
Since the epoch length is relatively short (typically a few minutes),
\NAME can construct indexes completely in memory 
and dump them to flash at the end of each epoch as a form of log structure.
It also facilitates support for time range queries (e.g., last 1 minute) by allowing query processing to use only a subset of the indexes
and makes eviction easy by removing the oldest set of indexes.

\subsection{Query Processing}
\label{sec:queryproc}

\NAME supports two query modes: offline and online.
In offline mode, \NAME handles queries while not recording any incoming traffic.
This mode is useful when rapid and quick query processing is crucial rather than accepting new packets.
On the other hand, when online, \NAME records new traffic data and processes queries using idle CPU cycles and I/O.
Online mode is useful when the user needs to query the archived data without disrupting recording.

Query processing in \NAME is made efficient by using indexes and data organization done in the recording stage.
A query handling task consists of flow lookup, header decompression, and packet reconstruction.
Each step can generate a list of flows, full headers, and full packets, respectively, based on the result from the previous step.
Depending on the query type, \NAME determines how further query processing it should proceed.

Flow lookup is the first stage of query processing.
This step uses indexes that can help refining query hits.
When given a time range in the query, \NAME chooses a subset of the indexes stored;
\NAME looks up selected indexes and intersects the lookup results (i.e., a set of flow locations) if there are any \emph{AND} operation.
Then, it verifies the full index keys stored together with compressed headers to finally refine the flow list.

Flow lookup is useful in answering simple and quick queries.
For example, existence test queries, which asks whether there is any flow seen in the traffic, can be done by using the flow lookup only.

The header decompression stage adds header content to the list of matching flows.
\NAME proceeds by decompressing the stored headers and emitting libpcap-compatible records.

The last query processing step is packet reconstruction, which builds and outputs full packets.
This stage uses the list of chunk locations attached to the compressed headers.
The specified chunks are read from disk, decompressed, and combined as a byte stream for a flow.
Then, using the payload length information in the headers, the byte stream is divided into packet payloads and appended to each header.
This stage is slower to process than the previous steps because reading chunks requires hard disk access, and decompressing chunks may cost a large amount of computation compared to the previous stages.

\subsection{Eviction}
\label{sec:eviction}

Eviction is simple in \NAME because it stores all traffic data in a log-structured way.
It can remove the oldest data without affecting the newer data.

\NAME currently does not send the evicted traffic data to an external longer-term archival service, which remains as future work.

\makeatletter{}\section{Implementation}
\label{sec:impl}

Our implementation of \NAME uses multiple CPU cores to provide high-speed recording and query processing while avoiding packet drops.

The main thread of \NAME accepts input packets from either a file or from standard input in the libpcap~\cite{www-tcpdump} trace format, which allows flexibility in choosing the packet acquisition method. This thread enqueues input packets to a buffer corresponding to their 5-tuple.
When it detects the end of a flow, it adds index entries for the flow in the flow indexes;
then, it dequeues and dispatches the flow to a worker thread.

The worker thread compresses the header and payload of the packets in the flow and stores the result to flash and disk.

The main thread also handles queries in online and offline mode.
Query processing uses continuations to implement online mode.
\NAME can stop query processing when it needs to continue archiving;
it can resume the query processing task from the last state where it left without losing the work done before.

\NAME is implemented in 17 K lines of C++ and is targeted to run on GNU/Linux x86-64.
For dictionary compression in both header and payload compression, \NAME uses LZO~\cite{www-lzo}, which makes a good compromise between compression ratios, compression speed, and decompression speed.

\makeatletter{}\section{Evaluation}
\label{sec:eval}

In this section, we evaluate three aspects of \NAME: (1) query performance; (2) recording throughput and memory use; and (3) storage use.

\subsection{Methodology}

Evaluating a traffic archiving system often requires full-packet network traffic.
This imposes two practical issues in privacy concerns and fair experimental comparisons.

Access to actual packet data must be handled carefully.
Both we and our IRB required that we avoid storing any traffic data possibly containing personally identifiable information (PII) on non-volatile storage, even if it is encrypted with a known encryption key.
This restriction means that payload processing had to be performed on the live data on-the-fly.

However, this live payload processing makes it difficult to fairly compare multiple system configurations.
Different experiment runs with a certain system setting will see different live traffic data, which may lead to completely different results.
If one experiment run uses multiple system configurations at a time, the system must handle more tasks than would have been required in a realistic situation with one system configuration.

To mitigate these difficulties,
we investigate both \emph{static} and \emph{dynamic} aspects of \NAME's performance.
In particular, we used the original live traffic with full payloads to investigate storage use under different system settings,
and we also used synthetic traffic based on anonymized header-only traces to examine system throughput and memory and storage use.

The following table shows the hardware we used for the evaluation.

{
\centering\small
\begin{tabular}{c||l}
	Component & Specification \\ \hline\hline
	CPU & 2x Intel Xeon E5-2450 at 2.1 GHz \\
	DRAM & DDR3 PC3-12800 48 GiB \\
	HDD (input files) & Western Digital RE4 500 GB \\
	HDD (output files) & 2x Seagate Barracuda LP 2 TB (RAID 0)\\
	SSD (output files) & Intel X25-M G2 80 GB \\
\end{tabular}
}

\begin{table*}
\centering\small
\begin{tabular}{c||c|c|c||c|c|c}
	Source & Description & Start Time (UTC) & Length & Original Bitrate & Packet Count & Avg Packet Size \\ \hline\hline
	\texttt{UNIV1} & University border router & 2011-07-28 19:17 & 7 days & 5.58 Mbps & 980 M & 430 B \\
	\texttt{UNIV2} & University border router & 2011-08-23 17:26 & 7 days & 4.08 Mbps & 739 M & 418 B \\
	\texttt{ISP*}  & CAIDA equinix-sanjose B  & 2012-11-15 13:00 & 30 minutes & 1.57 Gbps & 527 M & 669 B \\ 
\end{tabular}
\caption{Input workloads. The packet size excludes metadata (e.g., timestamps).}
\label{tbl:input-workloads}
\end{table*}

\subsection{Input Workloads}

We used two types of input workloads, \texttt{UNIV*} and \texttt{ISP*}, as summarized in Table~\ref{tbl:input-workloads}.

For \texttt{UNIV*} workloads, \NAME directly reads live traffic from a border router of a university.
Due to privacy concerns, \NAME performed the entire archival process on the live traffic without storing data with any PII on non-volatile storage devices.\footnote{The traffic collection and analysis was conducted with IRB approval for case \# \emph{anonymized for submission}.}

To measure the efficiency of payload compression, \NAME chunked, hashed, and compressed the traffic data using multiple chunking configurations (chunk size, variable- or fixed-length chunks).  It recorded the hashes and both original and compressed sizes.
Because performing this extra chunking and compression required significant system resources,
we restricted \NAME to process only the portion of the university border traffic that originates from or is destined for eight randomly sampled /24 subnets within the university network.  The specific identities of the subnets were kept anonymous from us.
Because traffic subsampling typically reduces redundancy within the input,
we believe that this measurement method makes our estimates of space savings from deduplication conservative.

The \texttt{ISP*} workload uses the CAIDA Anonymized Internet Traces 2012 Dataset~\cite{www-caida-traces-20121115}.
These traces contain headers only (i.e., timestamps, packet sizes, and anonymized packet headers).

This information is insufficient to provide enough data to evaluate \NAME because of its lack of payloads.
Therefore, we appended \emph{synthetic payloads} to each packet for two main purposes:
(1) as we can control the type of payload generation, we can evaluate \NAME from different angles with high reproducibility, which is often impossible with real traffic data;
and (2) by feeding artificial payload data into \NAME, we can evaluate \NAME's behavior under uncommon but important workloads, for example, anomalous network traffic or attacks targeted at \NAME itself.

In \texttt{ISP*} workloads, we used two types of payloads: redundant (\texttt{-RE}) and non-redundant (\texttt{-NR});
redundant payloads set each $i$-th payload byte to a product of $i$ and a random value specific to the packet,
whereas non-redundant payloads use completely random bytes for the payload.
\texttt{ISP-RE} shows how \NAME would behave with highly redundant traffic,
and \texttt{ISP-NR} often incurs the heaviest load on \NAME because it cannot reduce the amount of the total data it must handle through data compression.

Table~\ref{tbl:epoch-flow-params} shows common system parameters used for each input workload.
We used a longer maximum flow length for \texttt{UNIV*} because their relatively low bandwidth allows \NAME to store long flows for flow grouping,
whereas we limit it to 1 minute under \texttt{ISP*} workloads to avoid running out of system memory.

All results used variable-length chunking with a target chunking size of $4096$ bytes, if not specified;
\texttt{UNIV*} additionally used variable-length chunking with target average chunk size of $256$ and $1024$ bytes, and fixed-length chunking with target chunk size of $1024$ and $4096$ bytes, and no-chunking (treats the whole flow's payload as one chunk).

\begin{table}
\centering\small
\begin{tabular}{c||c|c}
	Input Workload  & Epoch Length & Maximum Flow Length \\ \hline\hline
	\texttt{UNIV1}  & 1 minute & 5 minutes \\
	\texttt{UNIV2}  & 1 minute & 5 minutes \\
	\texttt{ISP-RE} & 1 minute & 1 minute \\
	\texttt{ISP-NR} & 1 minute & 1 minute \\
\end{tabular}
\caption{The epoch length and the maximum flow length used in the experiments.}
\label{tbl:epoch-flow-params}
\end{table}

\subsection{Query Performance}

To demonstrate query performance, we used \texttt{ISP-NR}.  We omit the \texttt{ISP-RE} result because \texttt{ISP-NR} is more challenging traffic to record due to its data volume and thus we can expect conservative performance results.
All query outputs were written to \texttt{/dev/null} to prevent the measurement from being affected by the query output size.

\paragraph{Query types}
Query performance greatly varies by query type, e.g., match criteria, output data formats, query mode, and so on, as it performs different data structure access and computation to serve the queries.

We combine the following four characteristics to generate various query types in the query performance evaluation:
\begin{enumerate}
	\item \emph{Query mode} controls whether \NAME should stop recording while processing queries (``offline'') or should continue to record new traffic while handling query processing using idle system resources (``online'').
	\item \emph{Time range} defines what period of time \NAME should search for.  ``Last 1 min'' queries the traffic data recorded in the last 1 minute, while ``entire'' looks for all recorded traffic data.
	\item \emph{Retrieval format} specifies how much information the query wants.  ``Existence test'' queries only ask if there exists any flow or packet that matches query criteria.  ``Header'' and ``full'' queries request full headers and payloads for the matching packets.
	\item \emph{Query target} indicates how frequent hit the query will cause.  ``No hits'' means there will be no result at all matching the query criteria (e.g., \texttt{source address == 1.1.1.1}).  ``light hitters'' will have a few results in the recorded traffic (e.g., infrequently used ports), whereas ``heavy hitters'' will result in a large number of flows and packets as the query result (e.g., HTTP connections).
\end{enumerate}

\begin{table*}
\centering\small
\begin{tabular}{c|c||r}
	Time Range & Query Target & \textmu{}sec/query \\ \hline\hline
	Last 1 min & No hits       &    38.4 \\
	Last 1 min & Light hitters &    12.7 \\
	Last 1 min & Heavy hitters &    11.1 \\
	Entire     & No hits       & 2,740.9 \\
	Entire     & Light hitters &    15.1 \\
	Entire     & Heavy hitters &    13.2 \\
	\multicolumn{1}{c}{} \\
	\multicolumn{1}{c}{} \\
\end{tabular}\qquad
\begin{tabular}{c|c|c||r|r}
	Retrieval & Time Range & Query Target & Flows/sec & Packets/sec \\ \hline\hline
	Header & Last 1 min & Light hitters & 74,710.3 & 1,487,702.6 \\
	Header & Last 1 min & Heavy hitters & 99,322.6 & 1,594,282.1 \\
	Header & Entire     & Light hitters & 24,450.9 &   537,654.5 \\
	Header & Entire     & Heavy hitters & 84,704.6 & 1,563,395.2 \\ \hline
	Full   & Last 1 min & Light hitters &    112.6 &     1,601.0 \\
	Full   & Last 1 min & Heavy hitters &  5,819.3 &    98,434.1 \\
	Full   & Entire     & Light hitters &     93.6 &     1,870.3 \\
	Full   & Entire     & Heavy hitters &  5,814.9 &    98,384.2 \\
\end{tabular}
\caption{Query performance in offline mode on \texttt{ISP-NR}.  The left table shows the existence test results, the right table shows the packet retrieval results.}
\label{tbl:query-perf-offline}
\end{table*}

\begin{table*}
\centering\small
\begin{tabular}{c|c||r}
	Time Range & Query Target & \textmu{}sec/query \\ \hline\hline
	Last 1 min & No hits       &     87.9 \\
	Last 1 min & Light hitters &     42.4 \\
	Last 1 min & Heavy hitters &     37.0 \\
	Entire     & No hits       & 33,463.3 \\
	Entire     & Light hitters &     41.7 \\
	Entire     & Heavy hitters &     41.5 \\
	\multicolumn{1}{c}{} \\
	\multicolumn{1}{c}{} \\
\end{tabular}\qquad
\begin{tabular}{c|c|c||r|r}
	Retrieval & Time Range & Query Target & Flows/sec & Packets/sec \\ \hline\hline
	Header & Last 1 min & Light hitters & 19,053.4 & 441,648.4 \\
	Header & Last 1 min & Heavy hitters & 30,474.5 & 537,329.4 \\
	Header & Entire     & Light hitters &  4,401.8 &  89,276.5 \\
	Header & Entire     & Heavy hitters & 24,529.9 & 477,144.5 \\ \hline
	Full   & Last 1 min & Light hitters &     10.7 &     177.2 \\
	Full   & Last 1 min & Heavy hitters &    404.3 &   8,732.0 \\
	Full   & Entire     & Light hitters &     10.0 &     248.9 \\
	Full   & Entire     & Heavy hitters &  1,768.5 &  30,917.9 \\
\end{tabular}
\caption{Query performance in online mode on \texttt{ISP-NR}.  The left table shows the existence test results, the right table shows the packet retrieval results.}
\label{tbl:query-perf-online}
\end{table*}

\paragraph{Offline query mode}
Table~\ref{tbl:query-perf-offline} shows \NAME's performance in offline query mode.
With the query time range limited to the last 1 minute, 
even the slowest existence test query, which has no hits, thus making \NAME enumerate all candidate matches returned by the flow index, can be processed in $38.4$ \textmu{}s per query on average.
The query time jumps to $2.74$ ms per query when \NAME is instructed to investigate the entire time range.
Other query results involving light hitters or heavy hitters return in at most $15.1$ \textmu{}s on average
because \NAME often finds an early match and can stop further query processing.

\NAME can return a high number of flows and packets for queries requesting headers.
It can handle $24.5$ K flows per second, or $537.7$ K packets per second at minimum.
This high speed is due to the fact that \NAME can find matching packets quickly using flow indexes while making I/O to the fast flash drive only.

For retrieving full packets with payloads, \NAME shows about $0.1$--$5.8$ K flows per second, or $1.6$--$98.4$ K packets per second because \NAME accesses disk to read chunks to reconstruct full packet data.

In retrievals involving header or payload retrieval, heavy hitters tend to allow higher query performance because \NAME spends less time enumerating unused candidate flows.

When indexes cannot help query processing (e.g., using a TCP sequence number with no other query criteria), \NAME examined $97.3$ K flows per second, or $1,670.7$ K packets per second, similar to heavy hitters' throughput.

\paragraph{Comparison to query processing on pcap-format files}
To demonstrate how \NAME can make query processing interactive,
we performed existence test queries on \texttt{ISP*} \emph{header-only} pcap files with no compression, zlib, or LZO.
For the no-hit query type, queries on pcap required $179.4$, $257.5$, $174.2$ \textbf{seconds} per query, respectively for each compression method;
tcpdump's raw packet header scanning speed was at most $177.4$ K flows per second, or $3,024.3$ K packets per second, which was faster than \NAME's exhaustive packet header scanning without using indexes,
but when \NAME can use indexes, tcpdump is more than $4$ orders of magnitude slower than the worst case for \NAME requiring $2.74$ ms per query.
We expect that the performance gap will be even larger if tcpdump operates on full-packet traces because tcpdump must read full payloads to access packet headers.

This slow query processing with pcap files makes this approach inadequate in the situations where quick query processing is crucial.

\paragraph{Online query mode}
Table~\ref{tbl:query-perf-online} shows the query performance for the same types of queries when \NAME is actively recording new traffic data.
The average recording speed of \NAME is adjusted to $0.5$ Gbps so that \NAME has some idle time to process queries.

As the recording activity creates flash and disk activity, query throughput is slower during online operating than when offline.
However, most existence test queries finish in less than $100$ \textmu{}s on average, with the exception of no-hits queries over the entire time range, which took $33.5$ ms per query.
The reason why only this type of query exhibits high latency is because the epoch data and flow index are less likely to be cached in the in-memory system page cache because of the new data generated by recording;
\NAME must access indexes and header data on flash, leading to relatively longer query time.
Nevertheless, \NAME handles existence test queries at a high enough rate for users to make queries interactively.

If indexes are not used, \NAME could inspect $25.7$ K flows per second, or $471.0$ K packets per second, when also recording traffic.

\begin{table}
\centering\small
\begin{tabular}{c||c|c||r}
	Input & Chunk & Dictionary & Throughput \\
	Workload & Deduplication & Compression & (Mbps) \\ \hline\hline
	\texttt{ISP-RE} & No  & No  & 665.2 \\
	\texttt{ISP-RE} & Yes & No  & 700.4 \\
	\texttt{ISP-RE} & No  & Yes & \textbf{1,153.4} \\
	\texttt{ISP-RE} & Yes & Yes & 955.1 \\ \hline
	\texttt{ISP-NR} & No  & No  & \textbf{664.4} \\
	\texttt{ISP-NR} & Yes & No  & 629.4 \\
	\texttt{ISP-NR} & No  & Yes & 656.6 \\
	\texttt{ISP-NR} & Yes & Yes & 619.1 \\
\end{tabular}
\caption{Recording performance with different payload processing modes.}
\label{tbl:record-perf}
\end{table}

\subsection{Recording Performance}

Table~\ref{tbl:record-perf} presents recording performance with \texttt{ISP-RE} and \texttt{ISP-NR}.  To show system components' contribution to the recording performance, we turn on and off chunk deduplication and dictionary compression in payload compression as noted in the table.  We do not report the performance with \texttt{UNIV*} because we used multiple chunk size configurations simultaneously on the live traffic to avoid storing the original network traffic, and thus its performance was not representative.

With redundant payloads (\texttt{ISP-RE}), disabling chunk deduplication and enabling dictionary compression gives the best recording throughput of $1,153.4$ Mbps.
As we will see in Table~\ref{tbl:storage-use-payload}, \texttt{ISP-RE} greatly benefits from dictionary compression in reducing the data volume, and this leads to faster operation with less I/O.

When payloads contain no redundancy (\texttt{ISP-NR}), using dictionary compression and chunk deduplication only adds overhead as they require additional computation and I/O.

With chunk deduplication and dictionary compression enabled, recording throughput is at most $17.2$\% slower than the best throughput configuration.

Thus, the result with synthetic workloads suggests that \NAME can achieve high recording throughput with payload compression,
while it can further increase recording speed by adaptively bypassing chunk deduplication when its saving turns out to be less significant for the current traffic feed (e.g., anomalous or heavy attack traffic).

Throughout recording, the amount of DRAM consumed by \NAME did not exceed $15.7$ GB.

\subsection{Storage Space Use}

\begin{table}
\centering\small
\begin{tabular}{c||c|c|c|c}
	Input & Original & \multicolumn{3}{c}{Compressed Size (GB)} \\
	Workload & Size (GB) & \NAME & zlib & LZO \\ \hline\hline
	\multirow{2}{*}{\texttt{UNIV1}}  & 48.3 & \textbf{5.13} & -- & -- \\
	                                 &      & \textbf{(-89.4\%)} & & \\ \hline
	\multirow{2}{*}{\texttt{UNIV2}}  & 36.6 & \textbf{7.26} & -- & -- \\
	                                 &      & \textbf{(-80.2\%)} & & \\ \hline
	\multirow{2}{*}{\texttt{ISP-RE}} & 26.9 & \textbf{7.41} & 14.2 & 16.6 \\
	                                 &      & \textbf{(-72.5\%)} & (-47.2\%) & (-38.3\%) \\ \hline
	\multirow{2}{*}{\texttt{ISP-NR}} & 26.9 & \textbf{7.50} & 14.2 & 16.6 \\
	                                 &      & \textbf{(-72.1\%)} & (-47.2\%) & (-38.3\%) \\
\end{tabular}
\caption{Storage use by headers. The numbers in the parenthesis show the difference of the new chunk count and size from the original count and size.}
\label{tbl:storage-use-header}
\end{table}

\paragraph{Header} 
As shown in Table~\ref{tbl:storage-use-header},
\NAME effectively shrinks metadata and packet headers for all workloads.
It saves at least $80.2$\% of space in header information size for \texttt{UNIV1} and \texttt{UNIV2},
while it achieves $72.1$--$72.5$\% savings for the \texttt{ISP*} workloads;
this savings greatly exceeds the savings from simply compressing pcap-format traces with zlib and LZO, which required $14.2$ GB and $16.6$ GB, respectively, for \texttt{ISP*}.

The university trace headers are compressed more than the ISP inputs primarily because
\texttt{UNIV*} used longer flow length (5 minutes) than \texttt{ISP*} did (1 minute).
By having longer flows, \NAME's header compression becomes more effective, by not repeating a first full packet for each flow and having better efficiency for dictionary compression on the longer header data.

In addition, the university has fewer hosts talking to other hosts.
Common IP addresses thus appear more frequently, increasing the header compression efficiency.

Note that \texttt{ISP-RE} and \texttt{ISP-NR} have slightly different header sizes on storage because this header information includes the pointer to the payload chunk.

\begin{table*}
\centering\small
\begin{tabular}{c||c|c|c|c|c|c}
	Input & \multicolumn{2}{c|}{Original Chunk} & \multicolumn{3}{c|}{Unique Chunk} & Best-Cost Dedup Win \\
    Workload & Count & Size & Count & Size & Compressed Size & Compressed Size \\ \hline\hline
    \texttt{UNIV1} & 196.2 M & 371.5 GB & 184.9 M (-5.76\%) & 340.6 GB (-8.32\%) & 306.9 GB (-17.4\%) & 316.2 GB (-14.9\%)\\
	\texttt{UNIV2} & 127.0 M & 270.2 GB & 116.3 M (-8.43\%) & 239.2 GB (-11.5\%) & 218.8 GB (-19.0\%) & 227.9 GB (-15.7\%) \\
    \texttt{ISP-RE} & 57.9 M & 331.6 GB & 34.8 M (-39.9\%) & 307.9 GB (-7.15\%) & 51.8 GB (-84.4\%) & -- \\
    \texttt{ISP-NR} & 92.9 M & 331.6 GB & 92.9 M (-0.0\%) & 331.6 GB (-0.0\%) & 333.7 GB (+0.63\%) & -- \\
\end{tabular}
\caption{Payload compression results. The numbers in the parenthesis show the difference of the new chunk count and size from the original count and size.}
\label{tbl:storage-use-payload}
\end{table*}

\paragraph{Payload}
Table~\ref{tbl:storage-use-payload} shows the statistics for the original and unique chunks obtained by applying payload chunking and per-chunk dictionary compression.

For redundancy elimination for payloads, \NAME saves $8.32$--$11.5$\% for real world data in \texttt{UNIV1} and \texttt{UNIV2}.
After compression, the total savings increase to at most $17.4$--$19.0$\%.

While the compression ratios achieved for the synthetic payload ISP traces are obviously artificial, we discuss them briefly so that their effect on recording and query performance is clear.  In the redundant trace, \texttt{ISP-RE}, chunk deduplication reduces the number of chunks by $39.9$\%, but only saves $7.15$\% of total space: Deduplication was most effective for small chunks.  Dictionary compression, on the other hand, saved $84.4$\% of total space on this highly-compressible workload.

\paragraph{Optimizing payload compression for the best storage cost}
The storage used by payload data differs by how \NAME chunks and deduplicates payloads.
We vary the deduplication window size (how long the chunk index retains the stored chunk hash) with different chunking methods and examine the amount of redundancy \NAME detects and removes.

\begin{figure}
\centering
\includegraphics[width=0.90\columnwidth]{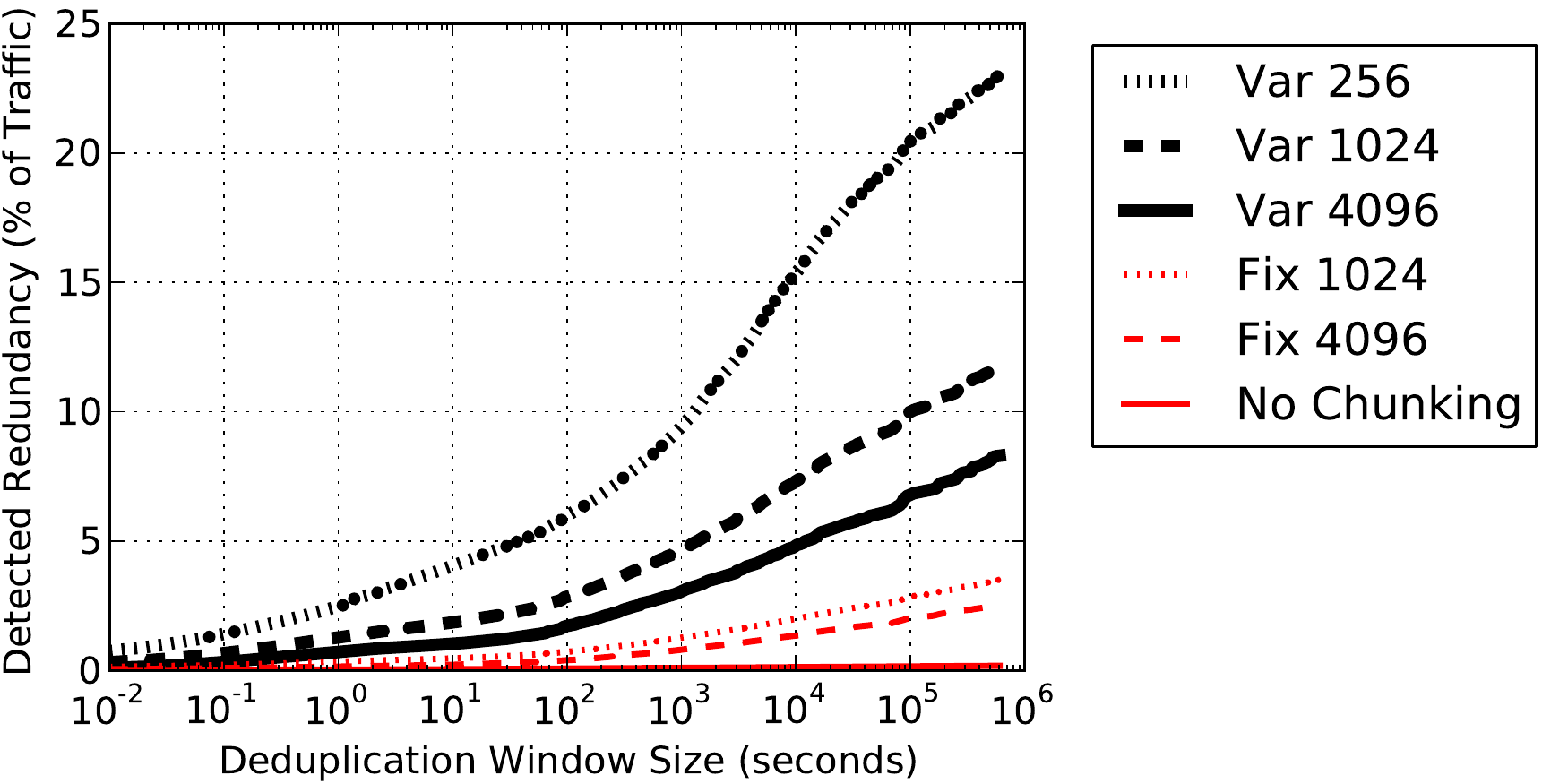}
\includegraphics[width=0.90\columnwidth]{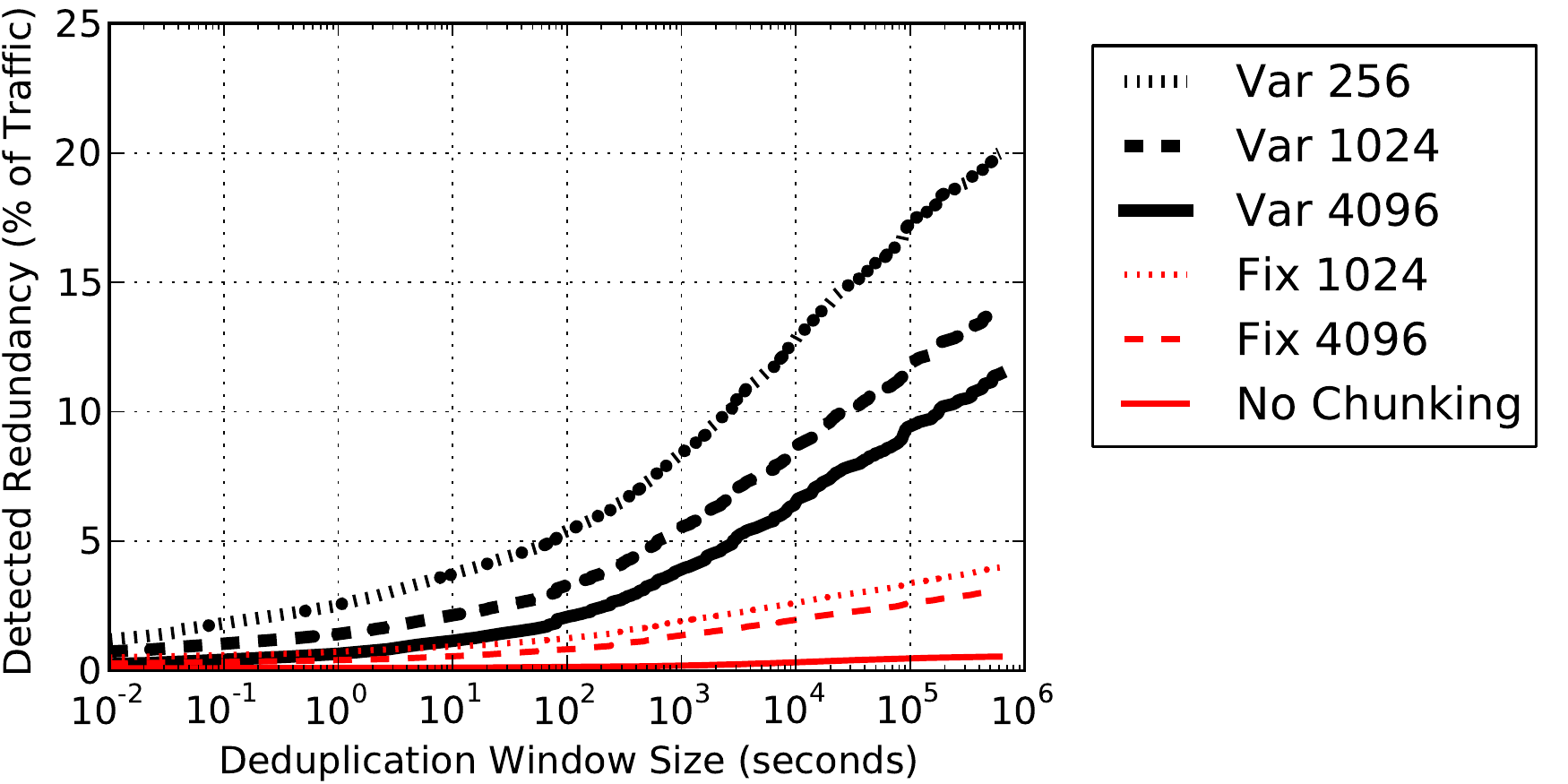}
\caption{Detected redundancy in payload from \texttt{UNIV1} (upper) and \texttt{UNIV2} (lower) when varying the deduplication window.  No per-chunk compression is applied.}
\label{fig:univ-redundancy-nocomp}
\end{figure}

\begin{figure}
\centering
\includegraphics[width=0.90\columnwidth]{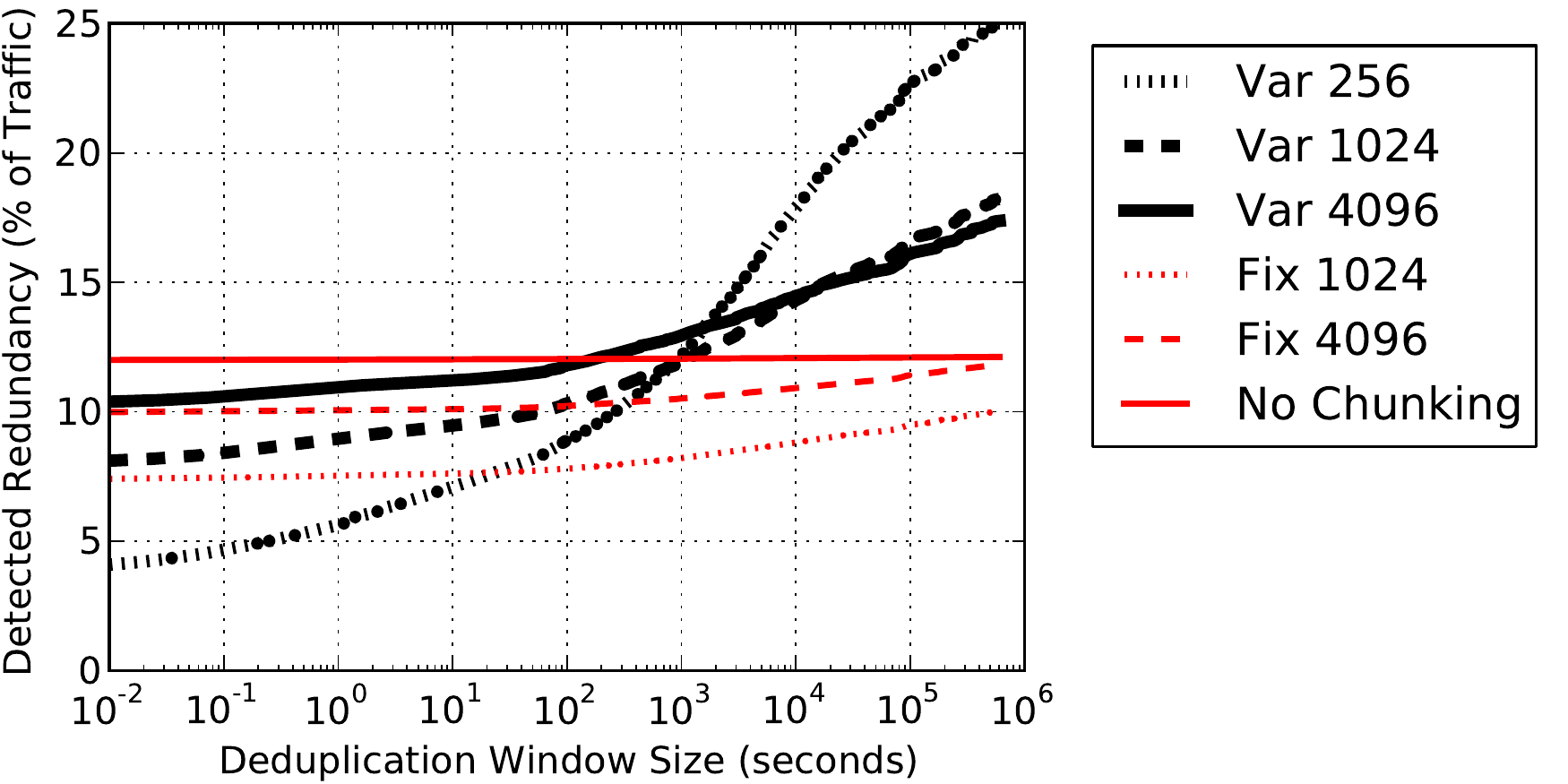}
\includegraphics[width=0.90\columnwidth]{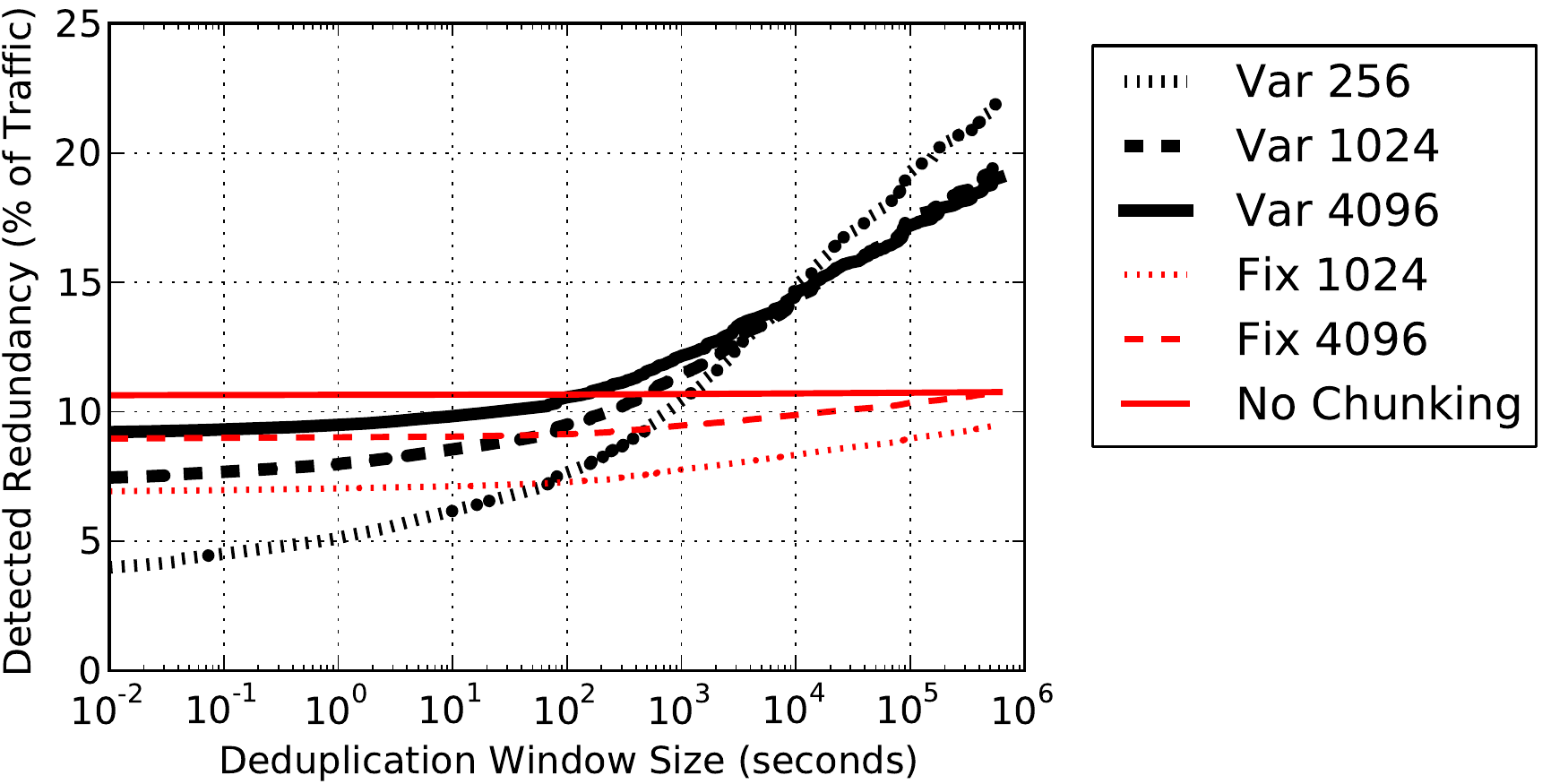}
\caption{Detected redundancy in payload from \texttt{UNIV1} (upper) and \texttt{UNIV2} (lower) when varying the deduplication window.  Per-chunk compression is applied.}
\label{fig:univ-redundancy-comp}
\end{figure}

Figure~\ref{fig:univ-redundancy-nocomp} and \ref{fig:univ-redundancy-comp} show how these factors affect the final space reduction, in uncompressed size (before applying dictionary compression) and compressed size (after applying dictionary compression; the final space use on storage).

Variable-length chunking allows more space savings than fixed-length chunking.
Except for a small deduplication window ($<$100 seconds) where fixed-length chunks can be compressed well with dictionary compression,
variable-length chunking helps find more duplicate chunks and results in better payload compression.

A larger deduplication window allows detecting a larger amount of redundant data.
However, the returns diminish:
beyond 10 K seconds, the slope of the detected redundancy decreases.

This diminishing effect leads to a sweet spot in storage cost savings.
If \NAME uses a large deduplication window, it saves HDD space, while using more SSD space to store the larger chunk index with more entries, and vice versa.

\begin{figure}
\centering
\includegraphics[width=0.90\columnwidth]{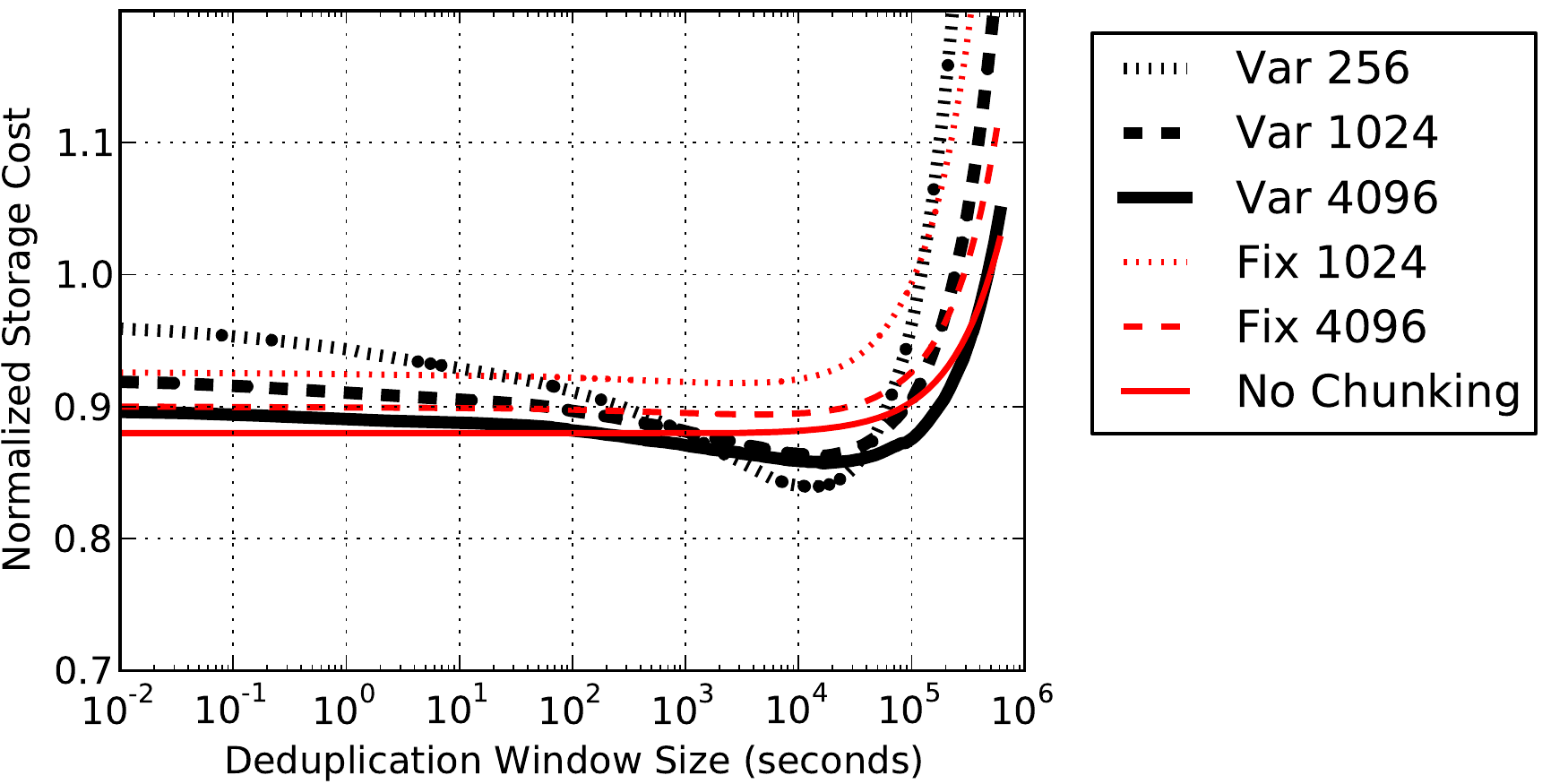}
\includegraphics[width=0.90\columnwidth]{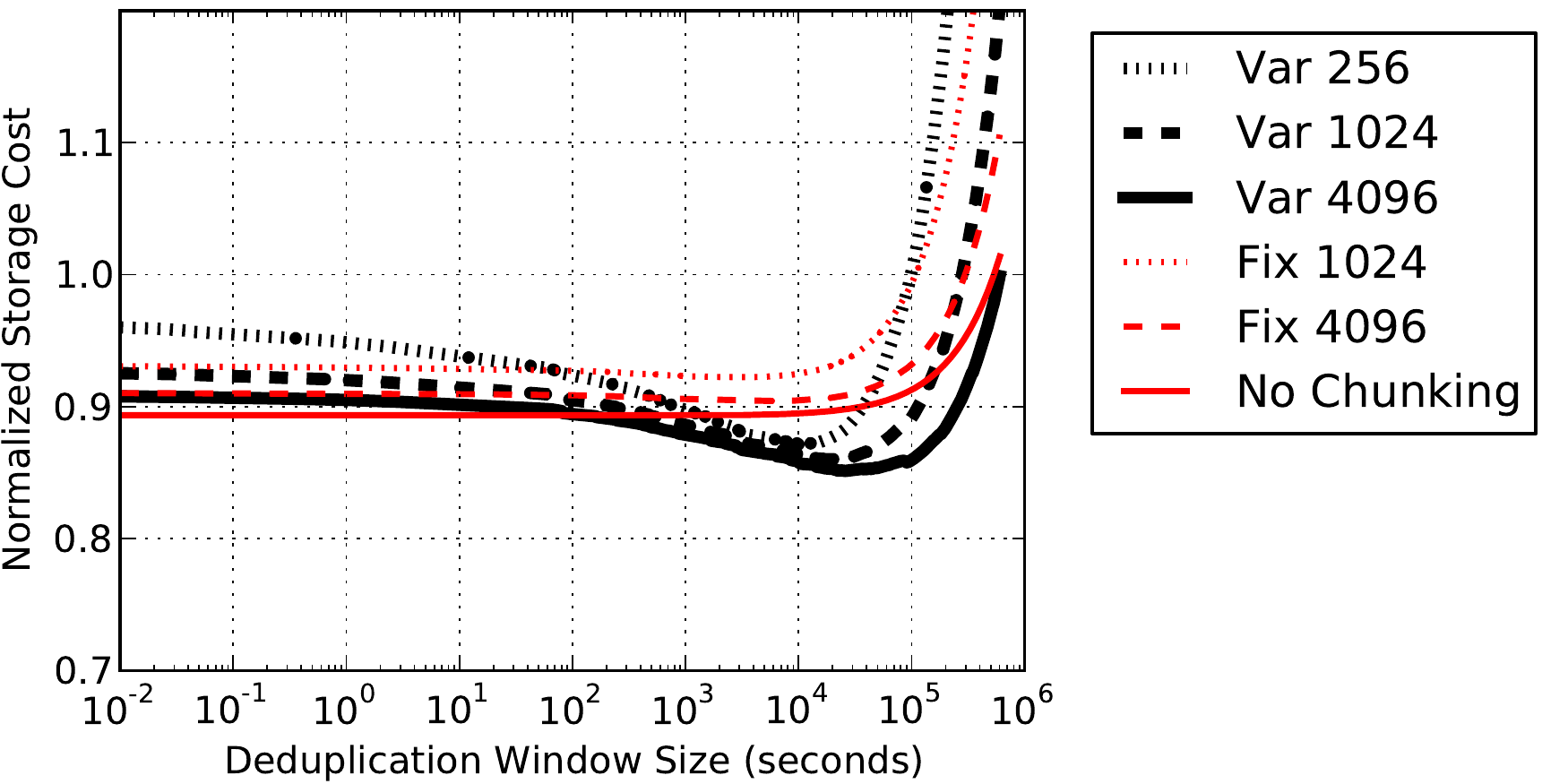}
\caption{Storage cost analysis for storing the payload data of \texttt{UNIV1} (upper) and \texttt{UNIV2} (lower) when varying the deduplication window.}
\label{fig:univ-cost}
\end{figure}

Figure~\ref{fig:univ-cost} plots how storage cost is affected by applying deduplication.
For this plot,
we set the HDD price to \$$0.0467$ per GB and the SSD price to \$$0.740$ per GB,
based on the market price of HDD and SSD as of January 2013.
The figure clearly shows that using variable-length chunking with a small target chunk size (e.g., $256$ bytes) can give the best cost effectiveness, while fixed-length chunking and no-chunking showed low redundancy.

However, a smaller chunk size has an extra cost: it makes many queries to the chunk index;
thus, the chunk size should be just small enough to fully utilize the flash drive where the chunk index is stored.

The optimal deduplication window size typically lies between $10$ K and $30$ K seconds;
a larger window increases the storage cost significantly due to too large chunk index size on flash.
Using the optimal deduplication window size for variable-length chunking with the 4096-byte average chunk size target, \NAME reduces storage cost by up to $14.9$\% (\texttt{UNIV1}) and $15.7$\% (\texttt{UNIV2}).

\begin{table*}
\centering\small
\begin{tabular}{c|c||c|c|c||c|c||c|c}
	Input & \multirow{2}{*}{Archival Method} & \multicolumn{3}{c||}{On-flash Size (GB)} & \multicolumn{2}{c||}{On-disk Size (GB)} & Total & Average \\
 Workload & & Header & Flow Index & Chunk Index & Header & Payload & Cost & of Two \\ \hline\hline
\texttt{UNIV1} & Original dump & -- & -- & -- & 48.3 & 371.5 & \$19.60 & \multirow{2}{*}{\$16.97} \\
\texttt{UNIV2} & Original dump & -- & -- & -- & 36.6 & 270.2 & \$14.33 & \\ \hline
\texttt{UNIV1} & LZO-compressed dump & -- & -- & -- & 29.8$^*$ & 326.9$^*$ & \$16.66 & \$14.52 \\
\texttt{UNIV2} & LZO-compressed dump & -- & -- & -- & 22.6$^*$ & 242.4$^*$ & \$12.38 & (-14.4\%) \\ \hline
\texttt{UNIV1} & \NAME & 5.13 & 1.59$^*$ & 0.141$^*$ & -- & 316.2 & \$19.84 & \$18.34 \\
\texttt{UNIV2} & \NAME & 7.26 & 0.976$^*$ & 0.139$^*$ & -- & 227.9 & \$16.84 & (+8.1\%)\\
\end{tabular}
\caption{Estimated total storage cost to archive 7-day traffic at \texttt{UNIV1} and \texttt{UNIV2}.  The numbers with a star ($^*$) are based on our predictions. The numbers in the parenthesis show the difference of the new storage cost from the storage cost with the original dump.}
\label{tbl:total-cost}
\end{table*}

\paragraph{Index}
uses an insignificant amount of flush storage.
For \texttt{ISP*} workloads,
\NAME used $199.3$ MB for the IP address index and $176.0$ MB for the port number index.
This translates to $12.14$ B per flow, or only $0.712$ B per packet.

\paragraph{Total storage cost}
Combined with the cost to store flow indexes and compressed headers on flash, the total storage cost using \NAME is comparable to the cost required by packet dump on hard disk, as summarized in Table~\ref{tbl:total-cost}.
In this table, flow index sizes are an estimated value based on the result with \texttt{ISP*}, and chunk index sizes are chosen based on Figure~\ref{fig:univ-cost}.
The header size in the compressed dump is based on our result with LZO-compressed \texttt{ISP*} traces while the payload size in the compressed dump is approximated using the detected redundancy using no-chunking in \texttt{UNIV*} (Figure~\ref{fig:univ-redundancy-comp}), $12.0$\% and $10.3$\%, respectively.
All other sizes are directly measured quantities.
We do not perform similar evaluation with \texttt{ISP*} workloads because they lack real payload data and are too short to give a meaningful result.

\paragraph{Summary}
\NAME provides high query performance and near-Gbps recording throughput while adding only $8.1$--$26.3$\% to the storage cost.

\makeatletter{}\section{Related Work}
\label{sec:related}

Time Machine~\cite{maier:sigcomm2008} is a traffic archiving system that exploits the heavy-tailed nature of flows to reduce the total data volume.
It keeps the first 10--20 KB of each connection and drops the rest; this allows retaining 91--96\% of all connections while saving 90\% of hard disk space.
They demonstrated recording about 4 days of traffic at a research institution using 2.1 TB hard disk space.
\NAME differs from Time Machine in that \NAME performs full traffic archival without loss, which protects \NAME from by attacks that exploit lossy archival.
In addition, \NAME exhibits high-performance query throughput, which requires only a few \textmu{}s per query; in contrast, Time Machine takes 125 ms per \emph{in-memory} query on average and may require minutes to complete disk-based queries due to its lack of efficient data layout and indexing and the use of hard disk only.
These differences in the completeness of archived data and the throughput of query processing make \NAME more useful in retrospective analysis and other security uses, but at substantially higher storage cost.

RRDtrace~\cite{Papadogiannakis:mascots2010} is another lossy traffic archiving system that focuses on storing a longer period of data on fixed storage space.
Different from Time Machine, RRDtrace applies more aggressive sampling on older data,
preserving more detail about recent data without requiring much storage space.
While RRDtrace is stronger than Time Machine against the attacks on the lossy archival because of the non-deterministic nature of its data reduction technique,
it neither exploits the redundancy in the network traffic to reduce the data volume 
nor provides query processing to quickly access the large amount of recorded traffic data.

Taylor \emph{et al.}~\cite{Taylor:atc2012} presents a network data storage system that shares several commonalities with \NAME.
In this work, they aggregate packets into connections and generate summary objects that describe the payloads of the connections as a list of key-value pairs, which are indexed by a set of partitioned indexes.
Unlike \NAME, however, this technique is an application protocol-specific approach because it requires interpreting payload data to generate summary objects with application-level domain knowledge.
Therefore, their system leaves a possibility for a bypass by using maliciously malformed payloads, encrypted connections (e.g., SSL), or the application protocols that are not being inspected.

NetFlow~\cite{www-netflow} and other aggregation-based monitoring techniques extract and record interesting features from the network traffic.
These features are typically very small compared to the original traffic volume, enabling their efficient storage for a long term.
However, it is hard to define features of interest for many security applications such as retrospective analysis because these feature are simply unknown in advance.
Further, since it discards the original traffic content, they are unsuitable for applications that require the original traffic for accurate evaluation of system performance and correctness.

NET-FLi~\cite{Fusco:2010vldbendow} and pcapIndex~\cite{Fusco:ccr2012} present a compact and fast indexing scheme that can be built on packet traces in the pcap format.
It creates a bitmap index whose bit indicates the presence of a certain value at the corresponding location of the original trace, then it applies a bitmap compression technique called COMPAX.
\NAME generally generates more compact indexes than these systems;
\NAME requires $0.356$ B/packet to index a field for ISP traces,
whereas COMPAX indexes require $3.64$ B/packet on average to index a field,
making \NAME's indexes an order of magnitude smaller than COMPAX indexes;
this is possible because \NAME performs flow-oriented data reorganization and allows its indexes to contain false positives that can be easily filtered out during query processing.

RasterZip~\cite{Fusco:imc2012} is a compression technique optimized for network traffic data, which performs column-wise compression opposed to byte-wise compression of conventional dictionary compressors.
While RasterZip can compress data in an agnostic way as long as the data items are well structured (e.g., a fixed number of columns),
\NAME's header compression can handle irregularities in the input data (e.g., IP only, TCP, and UDP headers).
In addition, RasterZip fails to remove redundancy that can be found with in-depth domain knowledge;
\NAME's header compression detects and removes intra-packet redundancy (e.g., valid IP checksum) because \NAME fully understands and can exploit the semantics of packet headers.

Selective Packet Paging (SPP)~\cite{Papadogiannakis:atc2012} strengthens packet capture systems from overload attacks that cause the systems to accept packets more than they can process.
\NAME is orthogonal to this work and can take advantage of it because SPP is used in the packet acquisition stage, which is outside of \NAME.

\makeatletter{}\section{Conclusion}
\label{sec:concl}

\NAME enables high-throughput recoding and query processing of network traffic without sacrificing the fidelity of the archived information or requiring high storage expenses.
In this paper, we show that \NAME can handle common types of queries within a few \textmu{}s on average with high-speed flow/packet retrieval, record near-Gbps full packet traffic, and use a small amount of storage space whose total hardware cost is comparable to hard disk-only solutions.
\NAME achieves these goals simultaneously by performing data reorganizations, using flash and disk carefully, and applying smart data compression, in a sophisticated and organized manner.

\balance

\setlength{\bibsep}{2pt}

\bibliographystyle{abbrvnat_noaddr} }{}

\end{document}